\DeclareSymbolFont{AMSb}{U}{msb}{m}{n}
\documentclass[reqno,centertags,10pt,a4paper,noamsfonts]{amsart}
\pdfoutput=1
\usepackage[svgnames]{xcolor}
\usepackage{graphicx}
\usepackage[english]{babel}
\usepackage[babel=true,expansion=alltext,protrusion=alltext-nott,final]{microtype}
\usepackage[margin={1.95cm},marginparwidth={1.95cm}]{geometry}
\usepackage[utf8]{inputenc}
\usepackage{textcomp}
\usepackage[sb]{libertine}
\usepackage[libertine,bigdelims,vvarbb]{newtxmath}
\useosf
\usepackage[varqu,varl]{inconsolata}
\usepackage[supsfam=LibertinusSerif-Sup,%
       supscaled=1.2,%
       raised=-.13em]{superiors}
\usepackage{mathtools}
\usepackage[T1]{fontenc}
\usepackage{textcomp}
\usepackage{amsthm}
\usepackage[inline]{enumitem}
\numberwithin{equation}{section}
\usepackage[normalem]{ulem}

\usepackage{float}  
\usepackage{caption}
\usepackage{subcaption} 
\usepackage{xspace}         
\usepackage[scientific-notation=true]{siunitx}      
\usepackage{pstricks}
\usepackage{tikz}
\usetikzlibrary{positioning} 
\usetikzlibrary{calc}
\usepackage{pgfplots}
\pgfplotsset{width=10cm,compat=1.9}
\usepackage{bm}
\usepackage{sidecap}
\usepackage{graphicx,color}

\usepackage{amsmath}
\DeclareFontFamily{U}{mathx}{}
\DeclareFontShape{U}{mathx}{m}{n}{<-> mathx10}{}
\DeclareSymbolFont{mathx}{U}{mathx}{m}{n}
\DeclareMathAccent{\widehat}{0}{mathx}{"70}
\DeclareMathAccent{\widecheck}{0}{mathx}{"71}

\providecommand{\mr}[1]{\href{http://www.ams.org/mathscinet-getitem?mr=#1}{MR~#1}}
\providecommand{\zbl}[1]{\href{https://zbmath.org/?q=an:#1}{Zbl~#1}}

\providecommand{\doi}[2]{\href{https://doi.org/#1}{#2}}

\newcommand{\C}{\mathcal{C}}

\definecolor{light_gray}{gray}{0.75}
\definecolor{lighter_gray}{gray}{0.5}
\colorlet{light_blue}{blue!20}
\definecolor{dark_green}{rgb}{0.0, 0.6, 0.0}
\definecolor{royal_blue}{rgb}{0.0, 0.22, 0.66}
\definecolor{salmon}{rgb}{1.0, 0.55, 0.41}
\definecolor{gold}{rgb}{0.8, 0.63, 0.21}
\definecolor{navy_blue}{rgb}{0.0, 0.0, 0.5}
\definecolor{crimson}{rgb}{0.79, 0.0, 0.09}
\definecolor{amethyst}{rgb}{0.6, 0.4, 0.8}
\definecolor{alizarin}{rgb}{0.82, 0.1, 0.26}
\definecolor{amaranth}{rgb}{0.9, 0.17, 0.31}
\definecolor{azure}{rgb}{0.0, 0.5, 1.0}
\definecolor{canaryyellow}{rgb}{0.82, 0.41, 0.12}
\definecolor{carrotorange}{rgb}{0.8, 0.33, 0.0}
\definecolor{cadmiumgreen}{rgb}{0.0, 0.42, 0.24}
\definecolor{copper}{rgb}{0.72, 0.45, 0.2}
\definecolor{aqua}{rgb}{0.5, 1.0, 0.83}
\definecolor{awesome}{rgb}{1.0, 0.13, 0.32}
\definecolor{candyapplered}{rgb}{1.0, 0.03, 0.0}
\definecolor{caribbeangreen}{rgb}{0.0, 0.8, 0.6}
\definecolor{indigo}{rgb}{0.0, 0.25, 0.42}

\DeclareMathOperator{\weaklystar}{\rightharpoonup\kern-2.2ex ^* \, \,}

\def\XXint#1#2#3{{\setbox0=\hbox{$#1{#2#3}{\int}$ }
\vcenter{\hbox{$#2#3$ }}\kern-.6\wd0}}

\newcommand{\R}{\mathbb R}
\newcommand{\N}{\mathbb N}
\newcommand{\Z}{\mathbb Z}
\renewcommand{\C}{\mathbb C}

\newcommand\norm[1]{\lVert #1 \rVert}

\newcommand\inner[1]{\langle #1 \rangle}

\newcommand\scpr{\boldsymbol{\cdot}}

\newcommand{\ra}{\rightarrow}

\newcommand{\mL}{\mathrm{L}}

\renewcommand{\phi}{\varphi}

\newcommand{\mH}{\mathrm{H}}

\newcommand{\T}{\mathbb{T}}

\newcommand{\ee}{\mathrm{e}}

\theoremstyle{plain}
\newtheorem{theorem}{Theorem}[section]

\newtheorem{corollary}[theorem]{Corollary}
\newtheorem{lemma}[theorem]{Lemma}

\newtheorem*{theorem*}{Theorem}

\theoremstyle{definition}
\newtheorem{definition}[theorem]{Definition}
\newtheorem{remark}[theorem]{Remark}
\newtheorem*{remark*}{Remark}

\usepackage[pagebackref=false]{hyperref}
\hypersetup{
	linktoc=all,
	bookmarksdepth=3,
	colorlinks=true,
	linkcolor=Green,citecolor=Orange,urlcolor=DarkBlue,
	pdftitle={Non-interacting v-reprensetability and Hohenberg-Kohn theorem on the 1D Torus},
	pdfauthor={Thiago Carvalho Corso}, 
	pdfsubject={},
	pdfkeywords={}} 
\begin{document}
\numberwithin{table}{section}
\title[Non-interacting v-reprensetability and Hohenberg-Kohn theorem on the 1D Torus]{v-representability and Hohenberg-Kohn theorem for non-interacting Schr\"odinger operators with distributional potentials in the one-dimensional torus}

\author[T.~Carvalho~Corso]{Thiago Carvalho Corso}
\address[T.~Carvalho Corso]{Institute of Applied Analysis and Numerical Simulation, University of Stuttgart, Pfaffenwaldring 57, 70569 Stuttgart, Germany}
\email{thiago.carvalho-corso@mathematik.uni-stuttgart.de}

\keywords{Density functional theory,  ensemble v-representability, pure state v-representability, Hohenberg-Kohn theorem, Kohn-Sham map, exchange-correlation potential, Schr\"odinger equation, distributional potentials}
\subjclass[2020]{Primary: 81Q10
 Secondary: 81V74, 34L40}

\date{\today}
\thanks{\emph{Funding information}:  DFG -- Project-ID 442047500 -- SFB 1481.  \\[1ex]
\textcopyright 2024 by the authors. Faithful reproduction of this article, in its entirety, by any means is permitted for noncommercial purposes.}
\begin{abstract}
In this paper, we show that the ground-state density of any non-interacting Schr\"odinger operator on the one-dimensional torus with potentials in a certain class of distributions is strictly positive. This result together with recent results from \cite{SPR+24} provides a complete characterization of the set of non-interacting v-representable densities on the torus. Moreover, we prove that, for said class of non-interacting Schr\"odinger operators with distributional potentials, the Hohenberg-Kohn theorem holds, i.e., the external potential is uniquely determined by the ground-state density. In particular, the density-to-potential Kohn-Sham map is single-valued, and the non-interacting Lieb functional is differentiable at every point in this space of $v$-representable densities. These results contribute to establishing a solid mathematical foundation for the Kohn-Sham scheme in this simplified setting.
\end{abstract}
\maketitle
\setcounter{secnumdepth}{2}

\section{Introduction}
    
    Density functional theory (DFT) has become a cornerstone of quantum chemistry and materials science; by reformulating the complex many-electron problem in terms of the electron density rather than the wavefunction, DFT offers a practical method to study the electronic structure of many-body quantum systems. A central piece to the outstanding success of DFT is the celebrated Kohn-Sham \cite{KS65} scheme, which seeks to reproduce the ground-state density of an interacting system of electrons via a fictitious system of non-interacting electrons. 
    
    However, the existence of such a fictitious non-interacting system is not well understood. This existence question is known as the v-representability problem and is a longstanding problem in the formulation of DFT. Despite its relevance (see, e.g. \cite{WAR+23,THS+22}), in the arguably most relevant case of continuous systems in three-dimensional space, a solution to the $v$-representability problem remains elusive. Nevertheless, in simplified cases such as lattice systems \cite{CCR85,PL21} and one-dimensional systems \cite{AS88,CS91,CS93,SPR+24} significant progress has been made. 
    
    Of special interest to us here is the recent paper by Sutter et al \cite{SPR+24}, where the authors established sufficient conditions for a density to be ensemble $v$-representable on the one-dimensional torus $
\T = \R/(2\pi \Z)$. More precisely, they showed that, for a fixed interaction potential $w$ satisfying suitable but rather general assumptions, any function $\rho :\T \rightarrow \R$ satisfying
    \begin{align}
        \sqrt{\rho} \in \mH^1(\T),\quad \int_\T \rho(x) \mathrm{d} x = N, \quad \mbox{and} \quad \rho(x) > 0 \quad \mbox{for all $x\in \T$,} \label{eq:necessary conditions}
    \end{align}
    can be realized as the density of a (possibly mixed) ground-state of a Hamiltonian of the form
    \begin{align*}
        H_N(v,w) = -\Delta + \sum_{i\neq j}^N w(r_i-r_j) + \sum_{j=1}^N v(r_i) \quad \mbox{acting on } \quad \mathcal{H}_N = \bigwedge^N \mL^2(\T),
    \end{align*}
    where $v$ is a (distributional) potential in $\mH^{-1}(\T)$. Here $\mH^1(\T)$ stands for the Sobolev space of square integrable functions with square integrable weak derivatives of first order, and $\mH^{-1}(\T)$ is the associated dual space (see Section~\ref{sec:notation} for precise definitions). 
    
    This result is rather remarkable as, to the best of the author's knowledge, it is the first rigorous sufficient criterion for both interacting and non-interacting $v$-representability in an infinite-dimensional and continuous system. However, as discussed in the conclusion of \cite{SPR+24}, many interesting questions regarding the $v$-representability problem on the one-dimensional torus remain unanswered; among them, the most important ones are perhaps the following: \begin{enumerate}[label=(\arabic*)]
    \item Is the condition~\eqref{eq:density space} also necessary, i.e., do any ground-state densities of $H_N(v,w)$ for arbitrary $v \in \mH^{-1}(\T)$ satisfy~\eqref{eq:density space}? 
    \item Is the class of potentials $v\in \mH^{-1}(\T)$ too large and merely a mathematical artifact, i.e., is there a smaller class of "more reasonable" potentials that suffices to represent all densities satisfying~\eqref{eq:density space}?
    \end{enumerate}
    
    In this paper, we answer these two questions in the case of non-interacting systems. More precisely, the main contribution of this work can be summarized as follows.
    \begin{enumerate}[label=(\roman*)]
    \item We show that, for non-interacting systems (i.e., $w=0$), the conditions in~\eqref{eq:necessary conditions} are not only sufficient but also necessary for $v$-representability. In particular, this gives a complete characterization of the set of non-interacting $v$-representable densities on $\T$. 
    \item We prove the Hohenberg-Kohn theorem for distributional potentials in the case of non-interacting systems, i.e., if $H_N(v,0)$ and $H_N(v',0)$, with $v,v'\in \mH^{-1}(\T)$, generate the same ground-state density, then $v=v'$ up to an additive constant. In particular, this implies that the Kohn-Sham density-to-potential map is a well-defined single-valued map from the space of densities satisfying~\eqref{eq:necessary conditions} to the space of potentials in $\mH^{-1}(\T)$.
    
    \item We show that the ground-state of any single-particle Schr\"odinger operator with potential in $\mH^{-1}(\T)$ is non-degenerate. In particular, for single-particle systems ($N=1$), the conditions in~\eqref{eq:necessary conditions} are in fact necessary and sufficient for \emph{pure-state $v$-representability} and the Kohn-Sham density-to-potential map is a well-defined \emph{smooth} and bijective map.
    
    \item On the other hand, we show that the second Hohenberg-Kohn theorem does not hold for excited states, i.e., there exists infinitely many (distinct up to an additive constant) potentials in $\mathbb{H}^{-1}(\T)$ whose Hamiltonian has the same excited state wave-function. 
    \end{enumerate}

\section{Main results}
 To state our main results precisely, let us first introduce some notation. 
 
 \subsection{Notation} \label{sec:notation} Throughout this paper, we let $\T = \R/(2\pi \Z)$ be the one-dimensional torus, and denote by $\mL^2(\T)$ the standard space of (equivalent classes) of measurable functions that are square integrable with respect to the Lebesgue measure, i.e., 
 \begin{align*}
     f\in \mL^2(\T) \quad \mbox{if and only if}\quad  \norm{f}_2^2 \coloneqq \inner{f,f} = \int_\T |f(x)|^2 \mathrm{d} x< \infty.
 \end{align*}
 Moreover, we denote by $\mathcal{H}_N$ the space of fermionic (or electronic) wave-functions on $\T^N$, i.e., the anti-symmetric tensor product space
 \begin{align*}
     \mathcal{H}_N \coloneqq \bigwedge^N \mL^2(\T). 
 \end{align*}
For a given wave-function $\Psi \in \mathcal{H}_N$, its (single-particle) density is defined as
 \begin{align*}
     \rho_{\Psi}(x) \coloneqq N \int_{\T^{N-1}} |\Psi(x,x_2,...,x_N)|^2 \mathrm{d} x_2... \mathrm{d} x_N.
 \end{align*}
We let $\mH^1(\T)$ denote the classical Sobolev space of functions $f \in \mL^2(\T)$ with weak derivative $\nabla f \in \mL^2(\T)$ endowed with the standard Hilbert norm, and we denote by $\mH^{-1}(\T)$ the associated dual space of $\mH^1(\T)$, i.e., the space of continuous linear functionals $v: \mH^1(\T) \rightarrow \C$ endowed with the operator norm.
 
Let us also introduce the following space of densities and potentials. For any $N\in \N$, we define $\mathcal{D}_N$ as the space
 \begin{align}
     \mathcal{D}_N \coloneqq \left\{\rho: \T \rightarrow \R \mbox{ such that} \quad \sqrt{\rho} \in \mH^1(\T), \quad \int_{\T} \rho(x) \mathrm{d} x = N, \quad \mbox{and}\quad \rho > 0 \right\} \label{eq:density space}
 \end{align}
 and $\mathcal{V}$ as the space
 \begin{align}
     \mathcal{V}\coloneqq\{ v \in \mH^{-1}(\T) :\quad v(f) \in \R \quad \mbox{for any real-valued function $f \in \mH^1(\T)$} \}. \label{eq:potential space}
 \end{align}

 Moreover, for any $v\in \mathcal{V}$, we shall denote by $h(v)$ the self-adjoint realization of the operator
 \begin{align*}
     h(v) = -\Delta + v 
 \end{align*}
 given as a form-perturbation of the Laplacian on the torus (periodic Laplacian). More precisely, $h(v)$ is the unique semi-bounded self-adjoint operator associated to the sesquilinear form
 \begin{align}
     q_{h(v)} : \mH^1(\T) \times \mH^1(\T)\rightarrow \C, \quad q_{h(v)}(\phi,\psi) = \int_\T \overline{\nabla \phi(x)} \scpr \nabla \psi(x)  \mathrm{dx} + v(\overline{\phi} \psi).\label{eq:quadraticform}
 \end{align}
 For more details of this construction, we refer to Section~\ref{sec:background}. For $N \in \N$, we define the associated non-interacting $N$-particles Hamiltonian $H_N(v)$ as
 \begin{align}
     H_N(v) = \sum_{j=1}^N 1 \otimes ... \otimes \overbrace{h(v)}^{\mathclap{j^{th} position}} \otimes ... \otimes 1 \quad \mbox{acting on} \quad \mathcal{H}_N. \label{eq:N-body hamiltonian}
 \end{align}
 
\subsection{Main results}

The first theorem we present here concerns the non-degeneracy of the ground-state for single-particle Hamiltonians with potentials in $\mathcal{V}$. This result plays a key role in the proof of the subsequent results and can be stated as follows.
\begin{theorem}[Non-degenerate single-particle ground-state] \label{thm:non-degenerate}
Let $v \in \mathcal{V}$ and $h(v) = -\Delta +v$ be the single-particle operator defined in~\eqref{eq:quadraticform}, then the ground-state of $h(v)$ is non-degenerate, and the unique (up to a global phase) normalized ground-state wave-function $\phi_v \in \mH^1(\T)$ is strictly positive everywhere, i.e., there exists a constant $c = c(v)>0$ such that
\begin{align*}
    \phi_v(x) > c \quad \mbox{for any $x\in \T$.}
\end{align*}
\end{theorem}

\begin{remark}[Existence of ground-state] \label{rem:ground-state} Since the quadratic form domain of $h(v)$ is $\mH^1(\T)$, which is compactly embedded in $\mL^2(\T)$, the operator $h(v)$ has compact resolvent and therefore discrete spectrum. In particular, ground-state wave-functions of $h(v)$ and its $N$-particles version $H_N(v)$ are guaranteed to exist. 
\end{remark}

As a consequence of Theorem~\ref{thm:non-degenerate}, we obtain the following necessary conditions for non-interacting $\mathcal{V}$-representabilitiy\footnote{Here we use the term $\mathcal{V}$-representability instead of $v$-representability to emphasize the class of potentials under consideration.}. 
\begin{corollary}[Necessary conditions for non-interacting $\mathcal{V}$-representabilitiy] \label{cor:necessary conditions} Let $N \in \N$ and $H_N(v)$ denote the $N$-particles non-interacting Hamiltonian defined in~\eqref{eq:N-body hamiltonian}. Then the density of any mixed ground-state $\Gamma$, 
\begin{align*}
    \rho_\Gamma(x) = N \int_{\T^{N-1}} \Gamma(x,x_2,...,x_N,x,x_2,...,x_N) \mathrm{d} x_2 ... \mathrm{d} x_N, 
\end{align*}
satisfies
\begin{align*}
    \sqrt{\rho_\Gamma} \in \mH^1(\T), \quad  \int_\T \rho_\Gamma(x) \mathrm{d} x = N \quad \mbox{and}\quad \rho_\Gamma(x) > 0 \quad \mbox{for any $x \in \T$.}
\end{align*}
\end{corollary}
Combining the above result with \cite[Theorem 1]{SPR+24}, we obtain a complete characterization of the set of non-interacting $\mathcal{V}$-representable densities on $\T$.
\begin{theorem}[Characteriztion of $\mathcal{V}$-representable densities]\label{thm:representability} A function $\rho : \T \rightarrow \R$ is the ground-state density of a $N$-particles non-interacting Hamiltonian $H_N(v)$ for some potential $v\in \mathcal{V}$ if and only if $\rho \in \mathcal{D}_N$ with $\mathcal{D}_N$ defined according to~\eqref{eq:density space}.
\end{theorem}

The second natural question that was left open in \cite{SPR+24} is whether one can recover the potential $v$ from the ground-state density $\rho$. This question is not only natural but also specially relevant for (Kohn-Sham) DFT because it implies that the Kohn-Sham density-to-potential map $\rho \mapsto v^{\rm KS}(\rho)$ is single-valued. Moreover, as highlighted in \cite[Corollary 19]{SPR+24}, the existence of a unique potential $v$ is also directly connected to the differentiability of the convex Lieb functional, which plays an important role in approximate schemes. Our next result provides an affirmative answer to this inverse problem for the case of non-interacting systems. 

\begin{theorem}[Hohenberg-Kohn theorem]\label{thm:HK} Let $N\in \N$ and suppose that $\rho \in \mathcal{D}_N$ is a ground-state density of $H_N(v)$ and $H_N(v')$ for $v, v'\in \mathcal{V}$. Then $v$ and $v'$ are equal up to a constant, i.e.,  $v = v' +c$ for some $c\in \R$.
\end{theorem}

\begin{remark} Note that Theorem~\ref{thm:HK} is not a special case of previous results \cite{Lie83, Zho12, Gar18,Lam18,Gar19, LBP20} because the class of potentials investigated here is rather large and include distributions such as the Dirac delta distribution. In fact, our results do not rely on the usual unique continuation for the $N$-particles wave-function \cite{Geo79, SS80, Gar18, Kur97} as such results only guarantee that the wave-function does not vanish on a set of Lebesgue measure zero on $\T^N$, which may correspond to the support of delta-type distributions. \end{remark}

Theorem~\ref{thm:HK} guarantees that the $N$-particles Kohn-Sham (KS) density-to-potential map is a well-defined and single-valued map from $\mathcal{D}_N$ to the quotient space
\begin{align*}
    \mathcal{V}/\{1\} \coloneqq \{[v] : v,v' \in [v], \mbox{ if $v-v'=$ constant} \},
\end{align*}
i.e., if we identify potentials that differ only by an additive constant. Moreover, the existence of a ground-state (see Remark~\ref{rem:ground-state}) implies that the KS map $v^{\rm KS}_N : \mathcal{D}_N \rightarrow \mathcal{V}/\{1\}$ is also surjective. In particular, the fibers of $v^{\rm KS}_N$, which corresponds to the set of ensemble ground-state densities of $H_N(v)$,
\begin{align*}
    (v^{KS}_N)^{-1}(\{[v]\}) = \mathcal{D}^{\rm ens}_N(v) \coloneqq \left\{ \rho = \sum_{j=1}^m t_j \rho_{\Psi_j}: \quad 0\leq t_j \leq 1, \quad \sum_{j=1}^m t_j = 1,\quad \mbox{and}\quad \Psi_j \mbox{ ground-state of $H_N(v)$}\right\}, 
\end{align*}
are all non-empty and disjoint, hence form a partition of the set $\mathcal{D}_N$. From this observation, we conclude that the space of potentials $\mathcal{V}$ is not only sufficient but also necessary to represent all the densities in $\mathcal{D}_N$.

However, we note that, due to possible degeneracies of the ground-state, not every density in $\mathcal{D}_N$ might be pure-state $\mathcal{V}$-representable. For instance, if we can find a potential $v \in \mathcal{V}$ such that the set of pure ground-state densities of $H_N(v)$,
\begin{align*}
    \mathcal{D}_N^{\rm pure}(v) \coloneqq \{\rho_{\Psi} : \Psi \mbox{ ground-state of $H_N(v)$}\}
\end{align*}
is not convex, then 
\begin{align*}
    \mathrm{conv}(\mathcal{D}^{\mathrm{pure}}_N(v)) \setminus \mathcal{D}^{\mathrm{pure}}_N(v) = \mathcal{D}^{\rm ens}_N(v) \setminus \mathcal{D}^{\rm pure}_N(v) \neq \emptyset
\end{align*}
and any density in this set is not pure-state $\mathcal{V}$-representable. In fact, as the subsets $\mathcal{D}^{\rm ens}_N(v)$ form a partition of $\mathcal{D}_N$, the lack of convexity of $\mathcal{D}^{\rm pure}_N(v)$ for some $v\in \mathcal{V}$ is the only obstruction to \emph{pure-state} (non-interacting) $\mathcal{V}$-representability. 

In particular, as the set $\mathcal{D}_1^{\rm pure}(v)$ for any $v\in \mathcal{V}$ consists of a single density by Theorem~\ref{thm:non-degenerate}, we obtain a complete solution to the \emph{pure-state} $\mathcal{V}$-representability problem in the case of a single-particle ($N=1$). Moreover, the proof of this result does not rely on the results from \cite{SPR+24}.

\begin{theorem}[Pure-state $\mathcal{V}$-representability for $N=1$]\label{thm:HK single-particle} Every density in $\mathcal{D}_1$ is pure-state $\mathcal{V}$-representable by a unique (up to an additive constant) potential $v\in \mathcal{V}$. In particular, the unique Kohn-Sham density-to-potential map is given by
\begin{align}
    v^{\rm KS}_1 : \mathcal{D}_1 \rightarrow \mathcal{V}/\{1\}, \quad \quad \rho \mapsto v^{\rm KS}_1(\rho) = \frac{\Delta \sqrt{\rho}}{\sqrt{\rho}}. \label{eq:KSmap}
\end{align}
Moreover, this map is  smooth\footnote{Note that $\mathcal{D}_1$ is a smooth manifold, as it is an open set of the closed subspace $\{f\in \mH^1(\T;\R) : \int_{\T} f(x) \mathrm{d}x  =1 \}$ of the Banach space $\mH^1(\T;\R)$. Hence, differentiability for maps in $\mathcal{D}_1$ has a well-defined meaning.} and bijective.\end{theorem}

As a last result, we show that, while the second part of the Hohenberg-Kohn theorem (cf. \cite[HK2 Theorem]{PTC+23}) still holds for ground-state wave-functions in the non-interacting case, i.e., two different potentials with the same ground-state wave-function can only differ by an additive constant, the same is not true for excited-states. This is a drawback of extending the class of admissible potentials to include distributions as the Dirac delta.
\begin{theorem}[No Hohenberg-Kohn for excited states] \label{thm:excited}   For any $v\in \mathcal{V}$ and any real-valued excited state $\phi_k$ of $h(v)$ with $k\geq 2$, there  exist (uncountable many) potentials $v' \in \mathcal{V}$ such that $v-v'$ is not constant and $\phi_k$ is also an excited state of $h(v')$.
\end{theorem}

\subsection{Outline of the paper}

Let us now outline the key steps in the proof of our results, and how they are distributed in the next sections.

The proof of Theorem~\ref{thm:non-degenerate} consists of three steps and is carried out in Section~\ref{sec:non-degenerate}. In the first step, we use classical results from the book by Reed and Simon \cite{RS78} (cf. Theorems XVIII.43 and 45) to show that the ground-state of $h(v)$ is non-degenerate and strictly positive almost everywhere. In the second step, we apply Courant's nodal domain theorem to conclude that the ground-state can not vanish in more than one point. If we were dealing with an interval $I$ with Dirichlet boundary conditions, these two steps would be enough to prove Theorem~\ref{thm:non-degenerate} because $I\setminus\{x_0\}$ consists of two connected components. In the torus, however, this is no longer true and we need a third step. This last step uses a gluing argument to obtain a contradiction with Courant's nodal domain theorem. More precisely, we use a gluing argument to construct ground-state densities that vanish on finitely many points, provided that a ground-state density vanishing on a single-point exists. 

The proof of the Hohenberg-Kohn Theorem~\ref{thm:HK} is presented in Section~\ref{sec:HK}. This proof consists of two main steps. The first step is the standard Hohenberg-Kohn argument, which shows that, if $H_N(v)$ and $H_N(v')$ have the same ground-state density, they must have a mutual ground-state wave-function. After this step, the usual argument used in previous proofs of the Hohenberg-Kohn theorem \cite{HK64,Lie83, PTC+23, Gar18} consists in dividing the Schr\"odinger equation
\begin{align}
    \left(H_N(v)-H_N(v')\right) \Psi = 0 \label{eq:differenceSE}
\end{align}
by this mutual ground-state wave-function $\Psi$, which is possible in a (almost everywhere) pointwise sense by unique continuation results. In our setting, this is no longer possible\footnote{In fact, in the single-particle case, the division by the ground-state argument is still possible (in an operator sense) thanks to the strict positivity of the ground-state (cf. Theorem~\ref{thm:non-degenerate}) and the algebra property of $\mH^1(\T)$ (Lemma~\ref{lem:algebra}), see, e.g., the proof of Theorem~\ref{thm:HK single-particle}. For $N\geq 2$, however, this argument does not apply as we have no control over the zero set of the $N$-particles ground-state $\Psi$.}  because we are dealing with distributional potentials which are not pointwise defined and whose support may have Lebesgue measure zero. To overcome this difficulty, we appeal to the fact that $h(v)$ has discrete spectrum, and combine a spectral representation of the dual space $\mH^{-1}(\T)$ (Lemma~\ref{lem:dual rep}) with Theorem~\ref{thm:non-degenerate}. More precisely, we show that, if $\Psi$ satisfies~\eqref{eq:differenceSE} in a weak sense,  has finitely many natural orbitals, and one of these orbitals can be chosen strictly positive, then the difference of potentials $(v-v')$ is much more regular than expected, namely, belongs to $\mH^1(\T)$. Combining this extra regularity with some linear algebra arguments, we can then show that $(v-v')$ must be constant, which completes the proof of Theorem~\ref{thm:HK}.

The proofs of Theorems~\ref{thm:HK single-particle} and~\ref{thm:excited} are also presented in Section~\ref{sec:HK}. The former follows from the existence, strict positivity, and non-degeneracy of the ground-state, plus some fairly standard arguments to show smoothness of a map. The proof of the latter relies on the simple observation that any (real-valued) excited state $\phi_k$ must vanish at a point, and therefore, adding a Dirac's delta at that point still preserves $\phi_k$ as an excited state.

For the sake of completeness, we present the necessary mathematical background for our proofs in some detail in Section~\ref{sec:background}. These comprises the precise definitions and a few well-known properties of the Sobolev spaces $\mH^1(\T)$ and $\mH^{-1}(\T)$, the quadratic form construction of $h(v)$, and the definition and simple properties of the natural orbital decomposition of a wave-function. In Section~\ref{sec:conclusion} we conclude with a brief discussion on some possible extensions of our main results and some natural open questions.

\section{Mathematical background}
\label{sec:background}
In this section we briefly review the mathematical background necessary for the proofs of our main results.

\subsection{Sobolev and dual spaces on the torus}
We begin by recalling the definitions of $\mH^1(\T)$ and $\mH^{-1}(\T)$. 

\begin{definition}[Sobolev spaces] We denote by $\mH^1(\T)$ the closure of the space $C^\infty(\T)$\footnote{Alternatively, one can define $\mH^1(\T)$ as the closure of $C^1$ functions on the interval $(0,2\pi)$ that are continuous up to the boundary, have $\mL^2$ integrable derivative, and satisfy $f(0) = f(2\pi)$.} with respect to the norm
\begin{align}
    \norm{\phi}_{\mH^1}^2 \coloneqq \norm{\phi}_{\mL^2}^2 + \norm{\nabla \phi}_{\mL^2}^2.  \label{eq:Sobolev norm}
\end{align}
The space $\mH^{-1}(\T)$ is the set of continuous linear functionals on $\mH^1(\T)$ endowed with the operator norm
\begin{align}
    \norm{v}_{\mH^{-1}}=\sup_{f\in \mH^1(\T) \setminus \{0\}} \frac{|v(f)|}{\norm{f}_{\mH^1}}. \label{eq:dualnorm}
\end{align}
\end{definition}

As usual, we identify measurable functions in $f:\T \rightarrow \C$ with linear functionals on $\mH^1(\T)$ via the Riesz mapping
\begin{align}
    g\in\mH^1(\T) \mapsto f(g) = \inner{f,g} = \int_\T \overline{f(x)} g(x) \mathrm{d} x. \label{eq:Riesz}
\end{align}
That such a functional is well-defined and continuous for any function $f\in \mL^1(\T)$ is a consequence of H\"older's inequality and the Gagliardo-Nirenberg-Sobolev (GNS) inequality stated below (see~\eqref{eq:GNS inequality}). In fact, the GNS inequality implies that $\mH^1(\T)$ is an algebra of functions. More precisely, we have
\begin{lemma}[Algebra property of $\mH^1(\T)$] \label{lem:algebra}
Let $\phi, \psi \in \mH^1(\T)$, then $\phi \psi \in \mH^1(\T)$ and the following estimate holds
\begin{align}
    \norm{\phi \psi}_{\mH^1(\T)} \lesssim \norm{\phi}_{\mL^2}^{\frac12}\norm{\phi}_{\mH^1}^{\frac12} \norm{\psi}_{\mH^1} + \norm{\psi}_{\mL^2}^{\frac12}\norm{\psi}_{\mH^1}^{\frac12} \norm{\phi}_{\mH^1}. \label{eq:Sobolev estimate}
\end{align}
In particular, the operator of multiplication by $\phi$, $\psi \mapsto M_\phi(\psi) = \phi \psi$, is bounded in $\mH^1(\T)$. Moreover, if $|\phi(x)| >0$ for every $x\in \T$, then this operator is invertible with inverse given by $M_{1/\phi}$.
\end{lemma}

\begin{proof}
    The proof of~\eqref{eq:Sobolev estimate} is straigthforward from the product rule, H\"older's inequality, and the well-known Gagliardo-Nirenberg-Sobolev inequality, which shows that any $\phi \in \mH^1(\T)$ is continuous and satisfies the bound
    \begin{align}
        \norm{\phi}_{\mL^\infty(\T)} \lesssim \norm{\nabla \phi}_{\mL^2}^{\frac12} \norm{\phi}_{\mL^2}^{\frac12} + \norm{\phi}_{\mL^2} \leq 2 \norm{\phi}_{\mL^2}^{\frac12} \norm{\phi}_{\mH^1}^{\frac12}.\label{eq:GNS inequality}
    \end{align}
\end{proof}

The following result is a simple consequence of the above lemma and will be useful to show that the KS map for a single-particle in~\eqref{eq:KSmap} is smooth. 
\begin{lemma}[Differentiability of push-forward] \label{lem:differentiable} Let $g\in C^\infty(\R;\R)$, then the (nonlinear) push-forward map $g^\#:\mH^1(\T;\R) \rightarrow \mH^1(\T;\R)$ given by
\begin{align*}
    (g^\#\psi)(x) = g\left(\psi(x)\right),\quad x\in \T.
\end{align*}
is smooth in $\mH^1(\T;\R)$.
\end{lemma}

\begin{proof} Let $\mathcal{B}(\mH^1(\T))$ denote the Banach space of bounded linear operators from $\mH^1(\T)$ to $\mH^1(\T)$. Then by Lemma~\ref{lem:algebra}, the map $M : \mH^1(\T) \rightarrow \mathcal{B}(\mH^1(\T))$, given by
\begin{align}
    \psi \in \mH^1(\T) \rightarrow M_{\psi} \in \mathcal{B}(\mH^1(\T)) \quad \mbox{where} \quad M_\psi(\phi) = \psi \phi, \label{eq:Mmap}
\end{align}
is continuous. Since this map is also linear, it is smooth. 

Next, we claim that for any $g\in C^3(\R;\R)$, the map $g^\#$ belongs to $C^1(\mH^1(\T;\R),\mH^1(\T;\R))$. To see this, first note that, since $g$ is locally Lipschitz, the chain rule yields
\begin{align*}
    \nabla (g^\#\psi) = \dot{g}(\psi) \nabla \psi, \quad \mbox{and therefore},\quad  \norm{\nabla (g^\# \psi)}_2 \leq \norm{\dot{g}}_{\mL^\infty(-C,C)} \norm{\nabla \psi}_2,
\end{align*}
where $C= \norm{\psi}_{\mL^\infty} \lesssim \norm{\psi}_{\mH^1}$ and $\dot{g}$ denotes the derivative of $g$. Hence, $g^\#$ maps $\mH^1(\T;\R)$ to $\mH^1(\T;\R)$. Furthermore, from the mean value inequality we find that 
\begin{align*}
    \norm{\nabla\left( g^\# (\psi+\delta) - g^\#(\psi) - M_{\dot{g}^\#(\psi)}\delta\right)}_{2}^2 &= \int_\T |\dot{g}(\psi+\delta) \nabla (\psi + \delta) - \dot{g}(\psi) \nabla \psi - \ddot{g}(\psi) \delta \nabla \psi - \dot{g}(\psi) \nabla \delta|^2\mathrm{d} x \\
    &= \int_\T \left|\left(\dot{g}(\psi+\delta)- \dot{g}(\psi)\right) \nabla \delta +\left(\dot{g}(\psi+\delta) - \dot{g}(\psi) - \ddot{g}(\psi) \delta\right)\nabla \psi\right|^2 \mathrm{d} x\\
    &\leq \norm{\ddot{g}}_{\mL^\infty(-C,C)}^2 \norm{\delta}_{\mH^1}^4 +  \norm{g^{(3)}}_{\mL^\infty(-C,C)}^2 \norm{\psi}_{\mH^1}^2 \norm{\delta}_{\mH^1}^4,
\end{align*}
where $C = \sup_{t \in [0,1]} \norm{\psi + t \delta}_{\mL^\infty} \lesssim \norm{\psi}_{\mH^1} + \norm{\delta}_{\mH^1}$, and $g^{(3)}$ denotes the third derivative of $g$. Thus, $g^\#$ is indeed $C^1$ and the derivative at $\psi \in \mH^1(\T;\R)$ is 
\begin{align*}
    \mathrm{d}_\psi g^\#(\delta) = \dot{g}(\psi) \delta = M_{\dot{g}^\# \psi} \delta.
\end{align*}
As this map is a composition of the $M$ map in~\eqref{eq:Mmap} and the map $\dot{g}^\#$, we conclude that $g^\#\in \C^2$, provided that $g\in C^4$. A bootstrap (induction) argument then shows that $g^\# \in \C^\infty$ for $g\in C^\infty$, which concludes the proof.
\end{proof}

\begin{remark}[Lie group structure of density spaces]\label{rem:Lie group} The GNS inequality implies that the set 
\begin{align}
\mathcal{D} \coloneqq \{f \in \mH^1(\T;\R): f >0 \}  \label{eq:D set}
\end{align}
is an open subset of the Banach space $\mH^1(\T;\R)$. This observation plays a key role in the proof of the results of \cite{SPR+24}. Moreover, together with Lemma~\ref{lem:algebra}, this implies that $\mathcal{D}$ is a Lie group\footnote{For precise definitions of (infinite dimensional) manifolds and Lie groups, we refer, e.g., to  \cite{KM97}.} with respect to pointwise multiplication. In particular, $\mathcal{D}_N$ defined in~\eqref{eq:density space} is a submanifold of codimension $1$, and the tangent space at each point $f\in \mathcal{D}_N$ can be identified with the set
\begin{align*}
    T_f\mathcal{D}_N = \left\{ \delta \in \mH^1(\T;\R); \int_\T \delta(x) \mathrm{d} x = 0 \right\}.
\end{align*}
\end{remark}

\subsection{Quadratic form definition of Hamiltonian} As a consequence of estimate~\eqref{eq:Sobolev estimate} and Young's inequality, the following bound holds:
\begin{align}
    v(|\phi|^2) \leq \norm{v}_{\mH^{-1}} \norm{\phi}_{\mH^1}^{\frac32} \norm{\phi}^{\frac12} \leq \norm{v}_{\mH^{-1}}\left(\epsilon \norm{\phi}_{\mH^1}^2 + \frac{C}{\epsilon} \norm{\phi}_{\mL^2}^2 \right)\quad \mbox{for any $\epsilon>0$.} \label{eq:Kato est}
\end{align}
In particular, the quadratic form induced by any $v \in \mH^{-1}(\T)$ is $\Delta$-form bounded with relative bound $0$. Therefore, by the KLMN theorem \cite[Theorem X.17]{RS75}, the operator $h(v)$ is a well-defined self-adjoint operator with quadratic form 
\begin{align*}
   q_{h(v)}(\phi,\psi) = \int_{\T} \overline{\nabla \phi(x)}\nabla \psi(x) \mathrm{d} x + v(\overline{\phi} \psi), \quad \mbox{for any $\psi,\phi \in Q\left(h(v)\right) \coloneqq \mH^1(\T)$.}
\end{align*} 
(For more details on this application of the KLMN theorem, we refer to \cite{RS75,Her89,SPR+24}.)

An useful consequence of this construction is that the quadratic form domain of $h(v)$ is the Sobolev space $\mH^1(\T)$. As this space is compactly embedded in $\mL^2(\T)$ by the standard Sobolev embedding theorem, the resolvent operator of $h(v)$ is a bounded operator form $\mL^2(\T)$ to $\mH^1(\T)$. Consequently, the resolvent is a compact operator and, by the spectral theorem, the spectrum of $h(v)$ is purely discrete, i.e., there exists a non-decresing sequence $\{\lambda_k\}_{k \in \N}$ and an $\mL^2$-orthonormal basis $\{\phi_k\}_{k \in \N}$ such that
\begin{align*}
    h(v) \phi_k = \lambda_k \phi_k \quad \mbox{and}\quad \lambda_k\uparrow \infty.
\end{align*}

The above spectral decomposition of $h(v)$ will be useful to prove the Hohenberg-Kohn theorem~\ref{thm:HK} in Section~\ref{sec:HK}. To be more precise, there we shall use the following representation of distributions in $\mH^{-1}(\T)$.
\begin{lemma}[Spectral representation of dual Sobolev space] \label{lem:dual rep} Let $v\in \mH^{-1}(\T)$ and $\{\phi_k\}_{k \in \N}$ be an orthonormal basis of eigenfunctions of $h(v)$ with corresponding eigenvalues $\{\lambda_k\}_{k\geq 1}$. Then a distribution $f\in \mathcal{D}'(\T)$ belongs to $\mH^{-1}(\T)$ if and only if there exists a sequence $\{c_j\}_{j \in \N}$ such that
\begin{align}
    \sum_{j\in \N} (1-\lambda_1+\lambda_j)^{-1} |c_j|^2 < \infty \quad \mbox{and}\quad f(\phi) = \sum_{j=1}^\infty c_j \inner{\phi_j, \phi}, \quad \mbox{for any $\phi\in \mH^1(\T)$.} \label{eq:c condition}
\end{align}
Moreover, in this case $c_j = f(\phi_j)$.
\end{lemma}
\begin{proof}
    The proof is rather standard, but we sketch the arguments for convenience of the reader. First, note that any $\phi \in \mL^2(\T)$, can be written as $\phi = \sum_{j\geq 1} a_j \phi_j$  where $a_j \coloneqq \inner{\phi_j,\phi}$.    Since $h(v)$ has quadratic form domain $\mH^1(\T)$ (see also~\eqref{eq:Kato est}), we find that 
    \begin{align*}
        \norm{\phi}_{\mH^1(\T)}^2 \sim \inner{\phi, \left(h(v) -\lambda_1 + 1\right) \phi} = \sum_{j=1}^\infty |a_j|^2 (\lambda_j -\lambda_1 + 1),
    \end{align*}
    where $\sim$ denotes equivalence of norms. Hence, $\phi \in \mH^1(\T)$ if and only if $\sum |a_j|^2 (1-\lambda_1+\lambda_j)<\infty$. Therefore, any $f$ of the form~\eqref{eq:c condition} belongs to $\mH^{-1}(\T)$ by Cauchy-Schwarz, i.e.,
    \begin{align*}
        f(\phi) = \sum_{j=1}^\infty c_j a_j \leq \left(\sum_{j=1}^\infty \frac{|c_j|^2}{(1+\lambda_j-\lambda_1)}\right)^{\frac12} \left(\sum_{j\geq 1} |a_j|^2 (1+\lambda_j-\lambda_1)\right)^{\frac12} \lesssim \norm{\phi}_{\mH^1(\T)}.
    \end{align*}
    Conversely, if $f\in \mH^{-1}(\T)$, then we can define $c_j \coloneqq f(\phi_j)$, and conclude by using the simple fact that
    \begin{align*}
        \left|\sum_{j=1}^\infty c_j a_j\right| \lesssim \left(\sum_{j=1}^\infty |a_j|^2 (1+\lambda_j-\lambda_1)\right)^{\frac12},\quad \mbox{for any sequence $\{a_j\}_{j\in \N}$,}
    \end{align*}
    if and only if $\{c_j\}_{j\in \N}$ satisfies~\eqref{eq:c condition}.
\end{proof}

\subsection{Single-particle density matrix and natural orbitals}  Let us now recall the definition of the single-particle density matrix and its natural orbital decomposition. 
\begin{definition}[Single-particle density matrix]  Let $N \in \N$, then for any $\Psi\in \mathcal{H}_N$ we define its single-particle density matrix as the function
\begin{align*}
\gamma_{\Psi}(x,y) \coloneqq N \int_{\T^{N-1}} \overline{\Psi(x,x_2,...,x_N)} \Psi(y,x_2,...,x_N) \mathrm{d} x_2... \mathrm{d} x_N.
\end{align*}
\end{definition}

Alternatively, one can see the single-particle density matrix as the integral kernel of the operator
\begin{align*}
    \inner{\phi, \gamma_{\Psi} \psi} = \inner{a(\phi) \Psi, a(\psi) \Psi}_{\mathcal{H}_{N-1}}, \quad \mbox{for $\phi, \psi \in \mathcal{H}_1$,}
\end{align*}
where $a(\phi)$ denotes the usual fermionic annihilation operator,
\begin{align}
    a(\phi):\mathcal{H}_N \rightarrow \mathcal{H}_{N-1} , \quad \left(a(\phi) \Psi \right)(x_1,...,x_{N-1}) = \sqrt{N} \int_\T \overline{\phi(x)} \Psi(x,x_1,...,x_{N-1}) \mathrm{d} x. \label{eq:annihilation}
\end{align}
It is well-known that the single-particle density matrix of any $N$-particles wave-function is a bounded, positive and trace-class operator satisfying
\begin{align*}
    \norm{\gamma_{\Psi}}_{\mathcal{H}_1\rightarrow \mathcal{H}_1} = \norm{\Psi}_{\mathcal{H}_N}^2 \quad \mbox{and}\quad \mathrm{tr} \gamma_{\Psi} = N \norm{\Psi}_{\mathcal{H}_N}^2.
\end{align*}
Hence, $\gamma_{\Psi}$ is a compact self-adjoint operator on $\mathcal{H}_1$ and, by the spectral theorem, there exists an orthonormal family of eigenfunctions $\{\phi_j\}_{j\leq M}$ and eigenvalues $\{n_k\}_{k \leq M}$ such that
\begin{align}
    0 < n_k \leq \norm{\Psi}_{\mathcal{H}_N}^2, \quad \sum_{k=1}^M n_k = N \norm{\Psi}_{\mathcal{H}_N}^2, \quad \mbox{and}\quad \gamma_{\Psi}(x,y) = \sum_{k=1}^M n_k \overline{\phi_k(x)} \phi_k(y) . \label{eq:natural orbital decomposition}
\end{align}
Here $M \in \N \cup\{+\infty\}$, i.e., $M$ does not need to be finite. The eigenfunctions $\phi_k$ are called the natural orbitals of $\Psi$, and $n_k$ are called the associated (natural) occupation numbers.

\begin{remark}[Non-uniqueness of natural orbitals] \label{rem:natural orbital}
    Note that the natural orbitals are not uniquely defined, as any linear combination of the orbitals with the same occupation number is again a natural orbital.
\end{remark}

For our proofs, we shall need two properties of the natural orbitals. The first one is the general fact that the natural orbitals of eigenfunctions of non-interacting operators can be chosen as eigenfunctions of the single-particle operator.

\begin{lemma}[Natural orbital decomposition of non-interacting ground-state] \label{lem:ground-state rep} Let $h$ be a semibounded self-adjoint operator on $\mL^2(\T)$ with purely discrete spectrum. Let $N\in \N$, then for any ground-state $\Psi\in\mathcal{H}_N$ of the non-interacting $N$-particles Hamiltonian
\begin{align*}
   H_N \coloneqq  \sum_{j=1}^N 1 \otimes 1... \otimes\overbrace{h}^{{\mathclap{j^{th} position}}} \otimes 1 ... \otimes 1, \quad\mbox{acting on} \quad \mathcal{H}_N
\end{align*} there exists an orthonormal basis $\{\phi_j\}_{j\in\N}$ of eigenfunctions of $h$ ordered in non-decreasing order, i.e., $\inner{\phi_k, h ,\phi_k} \leq \inner{\phi_{k+1} h, \phi_{k+1}}$, and a finite number $M\in \N$ such that $\{\phi_j\}_{j\leq M}$ are the natural orbitals of $\Psi$.
\end{lemma}

\begin{proof}
    The proof follows from the fact that, since the spectrum of $h(v)$ is discrete (see previous subsection), there exists an orthonormal basis of eigenfunctions $\{\phi_j\}_{j \in \N}$, and therefore, the Slater determinants
    \begin{align*}
        \Phi_{j_1,...,j_N} \coloneqq a(\phi_{j_1})^\ast a(\phi_{j_2})^\ast ... a(\phi_{j_N})^\ast 1, \quad 1\leq j_1 < j_2<... < j_N \in \N,
    \end{align*}
    where $1\in \C = \mathcal{H}_0$ is the vacuum state and $a(\phi)^\ast:\mathcal{H}_{N-1} \rightarrow \mathcal{H}_{N}$ is the creation operator
    \begin{align}
        \left(a(\phi)^\ast\Psi\right)(x_1,...,x_N) = \frac{1}{\sqrt{N}} \sum_{i=1}^N (-1)^{1+i} \phi(x_i) \Psi(x_1,...,x_{i-1},x_{i+1},...,x_N), \label{eq:creation}
    \end{align}
    form an eigenbasis of $H_N(v)$ for the $N$-particles space $\mathcal{H}_N$.
\end{proof}

The second result we shall need is the following simple formula for the overlapping density of a wave-function with its single-excitations. 

\begin{lemma}[Overlapping density with single-excitations] \label{lem:overlap} Let $\Psi \in \mathcal{H}_N$ and $\{\phi_j\}_{j=1}^M$ be its natural orbitals. Then for $\Phi = a(\psi)^\ast a(\phi_k) \Psi$, where $\psi \perp \{\phi_j\}_{j \leq M}$ and $a(\psi)^\ast$ and $a(\phi_k)$ are the creation and annihilation operators defined in~\eqref{eq:creation} and~\eqref{eq:annihilation}, we have
\begin{align}
    \rho_{\Psi, \Phi}(x) = \int_{\T^{N-1}} \overline{\Psi(x,x_2,...,x_N)} \Phi(x,x_2,...,x_N) \mathrm{d} x_2...\mathrm{d} x_N = n_k \overline{\phi_k(x)} \psi(x), \label{eq:overlap}
\end{align}
where $n_k>0$ is the occupation number of $\phi_k$. 
\end{lemma}

\begin{proof}
    First note that since $\psi \perp \{\phi_j: j \leq M\}$ can be assumed to be normalized, we can extend $\{\phi_j\}_{j \leq M}$ to an orthornomal basis $\{\phi_j\}_{j \in \N}$ such that $\phi_{M+1} = \psi$. Now consider $v\in \mL^\infty(\T)$ and denote by $V_N$ the $N$-particle operator of multiplication by \begin{align*}
        V_N(x_1,...,x_N) = \sum_{j=1}^N v(x_j).
    \end{align*}
    Then from the second quantization representation
    \begin{align*}
        V_N = \sum_{i,j\geq 1} v_{ij} a_i^\ast a_j, \quad\mbox{where $a_i= a(\phi_i)$, $a_j^\ast = a(\phi_j)^\ast$, and $v_{ij} = \inner{\phi_i, v \phi_j}$},
    \end{align*}
    and the canonical anti-commutation relations (CAR)
    \begin{align*}
        a_j a_k = -a_k a_j, \quad \mbox{and}\quad a_k a_j^\ast = \delta_{kj}- a_j^\ast a_k, \quad \mbox{where $\delta_{ij}$ is the Kronecker delta,}
    \end{align*}
   we find that
    \begin{align*}
        \inner{\Psi, V_N a(\psi)^\ast a(\phi_k) \Psi} = \sum_{i,j} v_{ij} \inner{\Psi, a_i^\ast a_j a_{M+1}^\ast a_k \Psi} =\sum_{i} v_{i M+1} \inner{a_i\Psi, a_k\Psi} + \sum_{i,j} v_{ij} \inner{a_i a_{M+1} \Psi, a_j a_k \Psi}. \end{align*}
    We can now use that $a_k \Psi = 0$ for any $k> M$ (since $\phi_k \perp \{\phi_j\}_{j\leq M}$ which are the natural orbitals of $\Psi$) to obtain
    \begin{align}
        \inner{\Psi, V_N \Phi} = \sum_{i=1}^M v_{iM+1}\inner{\phi_i, \gamma_{\Psi} \phi_k } = n_k v_{kM+1} = \int_\T v(x) n_k \overline{\phi}_k(x) \psi(x) \mathrm{d} x. \label{eq:firstid}
    \end{align}
    On the other hand, we have
    \begin{align}
        \inner{\Psi, V_N \Phi} = \int_{\T} v(x) \rho_{\Psi, \Phi}(x) \mathrm{d} x. \label{eq:secondid}
    \end{align}
    As~\eqref{eq:firstid} and \eqref{eq:secondid} holds for any $v\in \mL^\infty(\T)$, we obtain~\eqref{eq:overlap}.
\end{proof}

\section{Non-degeneracy of the ground-state}\label{sec:non-degenerate}

In this section, our goal is to prove Theorem~\ref{thm:non-degenerate} and Corollary~\ref{cor:necessary conditions}. To this end, we shall need two additional lemmas. 

The first lemma shows that the ground-state of $h(v)$ is non-degenerate and almost everywhere strictly positive. This kind of result is sometimes called Perron–Frobenius theorem; it can be understood as a unique continuation property but applies only to the ground-state of the single-particle system.

\begin{lemma}[Non-degenerate ground-state] \label{lem:non-degenerate} Let $v \in \mathbb{H}^{-1}(\T)$, then the ground-state of $h(v)$ is non-degenerate and (up to a global phase) satisfies 
\begin{align*}
    \phi_v(x) >0 \quad \mbox{for almost every $x\in \T$.}
\end{align*}
\end{lemma}

\begin{proof}
    For the proof, we shall use two results from the book by Reed and Simon \cite{RS78}. To simplify the presentation, we combine the statement of these two results in a single lemma.
\begin{lemma}[Theorem XIII.43 and XIII.45] \label{lem:RS} Let $h_0$ and $h$ be two semibounded self-adjoint operators in $\mL^2(\Omega,\mu)$, where $(\Omega,\mu)$ is a sigma-finite measure space. Suppose that 
\begin{enumerate}[label=(\roman*)]
\item \label{it:gs} $h_0$ has a simple and almost everywhere strictly positive ground-state.
\item \label{it:positivity} $\ee^{-th_0}$ is positivity preserving, i.e., it maps non-negative functions to non-negative functions.
\end{enumerate} 
Then, if there exists a sequence of bounded functions $\{v_n\}_{n\in\N}\subset \mL^\infty(\Omega,\mu)$ such that $h_0 + v_n$ converges in the strong resolvent sense to $h$, then the ground-state of $h$ (if existing) is also simple and strictly positive almost everywhere.
\end{lemma}

Thus, in order to prove Lemma~\ref{lem:non-degenerate}, it suffices to find an operator $h_0$ satisfying~\ref{it:gs} and~\ref{it:positivity} and a sequence of bounded functions $v_n$ such that $h_0 + v_n$ converges in the strong resolvent sense to $h(v)$. As usual in the applications of such results, we choose $h_0$ as the free (periodic) Laplacian.

So first, note that the ground-state of $h_0 = -\Delta$ is the constant function, and therefore strictly positive; thus assumption~\ref{it:gs} holds for $h_0$. Next, recall that the heat propagator of the periodic Laplacian is given by
    \begin{align*}
        (\ee^{-t h_0} f)(x) = (p_t \ast f)(x) = \int_0^{2\pi} p_t(x-y) f(y) \mathrm{d} y,
    \end{align*}
    where $p_t$ is the periodization of the heat kernel of the Laplacian in $\R$. As the latter is nothing but the standard Gaussian, we find that
    \begin{align*}
        p_t(x) = \sum_{k \in\Z} \frac{1}{\sqrt{4\pi t}} \ee^{-\frac{(x+2\pi k)^2}{4t}} \geq 0 \quad \mbox{for any $x\in [0,2\pi]$.}
    \end{align*}
    As convolution with a non-negative function is a positivity preserving operator, assumption~\ref{it:positivity} also holds. 

    To conclude, note that, since $\mL^2(\T)$ is dense\footnote{Recall that the inclusion $\mL^2(\T) \subset \mH^{-1}(\T)$ is given by the Riesz mapping introduced in~\eqref{eq:Riesz}.} in $\mH^{-1}(\T)$ (by Lemma~\ref{lem:dual rep}) and $\mL^\infty(\T)$ is dense in $\mL^2(\T)$, there exists a sequence of bounded functions $v_n \in \mL^\infty(\T)$ such that 
    \begin{align*}
        \lim_{n \ra \infty} \norm{v-v_n}_{\mH^{-1}(\T)} = 0.
    \end{align*}
    Consequently, by estimate~\eqref{eq:Sobolev estimate} in Lemma~\ref{lem:algebra},
    \begin{align*}
        |\inner{\phi, h(v) \psi} - \inner{\phi, h_0+v_n, \psi}| = |\inner{\phi, (v-v_n) \psi}| = |(v-v_n)(\overline{\phi} \psi)| \lesssim \norm{v-v_n}_{\mH^{-1}}\norm{\psi}_{\mH^1} \norm{\phi}_{\mH^1}.
    \end{align*}
    Therefore, $h_0 + v_n$ converges to $h(v)$ in the norm topology of operators in $\mathcal{B}\left(\mH^1(\T), \mH^{-1}(\T)\right)$. Using the first resolvent formula, we then conclude that $h_0 + v_n$ converges to $h(v)$ in the strong resolvent sense (see, e.g., \cite[Theorem VIII.25.(c)]{RS80} for a similar argument). Therefore Lemma~\ref{lem:RS} applies and the proof is complete.
\end{proof}

The previous lemma only guarantees that the ground-state $\phi_v$ does not vanish in a set of measure zero. To obtain the stronger strict positivity everywhere, we shall use the following version of the classical Courant nodal domain theorem \cite{CH89}. The proof of this result is exactly the same as in the case of Schr\"odinger operators with standard multiplication potentials. Yet, for the sake of completeness, we briefly sketch the main steps below.

\begin{lemma}[Courant nodal domain theorem] \label{lem:Courant} Let $\phi \in \mH^1(\T)$ be an eigenfunction of $h(v)$ with eigenvalue $\lambda$, then the number of nodal domains of $\phi$, i.e., the number of connected components of the (open) set $\{x\in \T: \phi(x) \neq 0\}$ is less or equal than $n(\lambda) \coloneqq \sum_{\mu \leq \lambda} \dim \ker\left( h(v)-\mu\right)$.
\end{lemma}

\begin{proof}
    First, let $\phi \in \mH^1(\T)$ be an eigenstate of $h(v)$ with eigenvalue $\lambda$. Let $\{O_j\}_{j \leq M}$ denote the connected components of $\{\phi\neq 0\}$, and define $\phi_j \coloneqq \phi\rvert_{O_j}$. Since $\phi$ is continuous by the GNS inequality~\eqref{eq:GNS inequality}, each of the sets $O_j$ is an open interval. Moreover, as $\phi = 0$ at the ends of $O_j$, the extension of $\phi_j$ by zero to the whole torus $\T$ belongs to $\mH^1(\T)$. Since $\overline{\phi_j} \phi_k = 0$ everywhere for $j\neq k$, we have
    \begin{align*}
        q_{h(v)}(\phi_j,\phi_k) = \int_{\T} \overline{\nabla \phi_j(x)} \nabla \phi_k(x)\mathrm{d} x + v(\overline{\phi_j} \phi_k) = \int_{\T} \overline{\nabla \phi_j(x)} \nabla \phi_k(x) = 0 \quad \mbox{if $j\neq k$.}
    \end{align*}
    In particular, as $\phi=\sum \phi_k$ is an eigenfunction of $h(v)$ with eigenvalue $\lambda$ we have
    \begin{align*}
        \lambda \norm{\phi_j}_{\mL^2}^2 = \lambda\inner{\phi,\phi_j}_{\mathcal{H}_1} = \inner{h(v) \phi, \phi_j}_{\mathcal{H}_1} = \int_{\T} \overline{\nabla \phi(x)} \nabla \phi_j(x) \mathrm{d} x + v(\overline{\phi} \phi_j) = \int_{\T} |\nabla \phi_j(x)|^2 \mathrm{d} x + v(|\phi_j|^2).
    \end{align*}
    Therefore, for any $\psi \in \mathrm{span}\{\phi_1,...,\phi_M\}$, i.e., $\psi = \sum_{j=1}^M c_j \phi_j$ for some $\{c_j\}_{j\leq M} \subset \C$, we have
    \begin{align*}
        q_{h(v)}\left(\psi,\psi\right) = \sum_{j=1}^M |c_j|^2 q_{h(v)}(\phi_j,\phi_j) = \lambda \sum_{j=1}^M |c_j|^2 \norm{\phi_j}_{\mL^2}^2 = \lambda \norm{\psi}_{\mL^2}^2. 
    \end{align*}
    As $\dim \mathrm{span}\{\phi_1,...,\phi_M\} = M$, the result now  follows from the (Courant-Fischer-Weyl) min-max principle.
\end{proof}
We are now ready to prove Theorem~\ref{thm:non-degenerate}.
\begin{proof}[Proof of Theorem~\ref{thm:non-degenerate}]
By Lemma~\ref{lem:non-degenerate} and Lemma~\ref{lem:Courant}, we see that $\phi_v(x)$ can vanish on at most one point in $\T$. Indeed, if there were at least two points where $\phi_v(x) = 0$, there would be at least two connected components of $\{x  \in \T: \phi_v(x) >0\}$ which contradicts the non-degeneracy of the ground-state by Lemma~\ref{lem:Courant}. However, to complete the proof we need to show that $\phi_v(x)$ can not vanish at a single point. 

To this end, we shall argue by contradiction. More precisely, we first assume that there exists a ground-state vanishing at a single-point, and then, via a gluing argument, we construct a ground-state of a different Hamiltonian that vanishes at two distinct points. By the preceding paragraph, this gives a contradiction and suffices to complete the proof.

So suppose that there exists a ground-state $\phi_v\geq 0$ of $h(v)$ for some $v\in \mathbb{H}^{-1}(\T)$ such that $\phi_v(x) = 0$ only at $x=0$. Without loss of generality, we also assume that the ground-state energy is $0$. Then, we define
\begin{align*}
    \widetilde{\phi}(x) \coloneqq \phi_v(x) \eta(x) + \eta(x-\pi/2) + \phi_v(x-\pi) \eta(x-\pi) + \eta(x-3\pi/2 )
\end{align*}
where $\eta(x) = \eta(x\,\mathrm{mod}\,2\pi)$ for some $\eta \in \C_c^\infty((-\pi/2,\pi/2);\R)$ satisfying 
\begin{align*}
   \eta(x) = 0 \quad \mbox{for $|x| \geq \pi/3$}, \quad 
   \eta(x) >0 \quad \mbox{for $|x| < \pi/3$,} \quad \mbox{and} \quad 
   \eta(x) = 1, \quad \mbox{for $|x|\leq \pi/8$.}
\end{align*}
Here we denote by $x\,\mathrm{mod} \, 2\pi$ the unique $y\in(-\pi,\pi]$ such that $y-x \in 2\pi \Z$. Then, clearly, $\widetilde{\phi} \in \mH^1(\T)$. Moreover, since the support of $\eta$ is contained in $[-\pi/3, \pi/3]$,  we have
\begin{align}
    \begin{dcases} \widetilde{\phi}(x) = \phi_v(x) \quad &\mbox{for $|x \,\mathrm{mod}\, 2\pi| \leq \pi/8$} ,\\
    \widetilde{\phi}(x) = \phi_v(x-\pi), \quad &\mbox{for $|x-\pi|\leq \pi/8$,}\\
    \widetilde{\phi}(x) > 0, \quad \mbox{otherwise}. \end{dcases} \label{eq:local}
\end{align}
In particular, $\widetilde{\phi}_v(x) \geq 0$ and $\widetilde{\phi}_v(x) = 0$ if and only if $x=0$ or $x=\pi$.

Next, let $\{p_j\}_{j=1}^3$ be a partition of the unity\footnote{That such partition of unity exists is a well-known fact in differential geometry. In the simple case of a torus they can be constructed explicitly from any function $\eta$ as above.} subordinate to the following open cover of $\T$:
\begin{align*}
    I_1 \coloneqq (-\pi/8,\pi/8) \,\mathrm{mod}\, 2\pi, \quad  I_2 \coloneqq (7\pi/8,9\pi/8)\quad \mbox{and}\quad  I_3 \coloneqq \{x \in \T: |x \, \mathrm{mod}\, 2\pi| > \pi/16, \quad |x-\pi| > \pi/16\}.
\end{align*}
In other words, each of the $p_j$ satisfies $p_j \in C^\infty(\T)$, $p_j(x) = 0$ for $x\not\in I_j$, $p_j(x) \geq 0$, and together they satisfy 
\begin{align}
    \sum_{j=1}^3 p_j(x) = 1, \quad \mbox{for any $x\in \T$.} \label{eq:partition unity}
\end{align}
Then, we define the distribution $\widetilde{v} \in \mathcal{D}'(\T)$ as
\begin{align*}
    \widetilde{v}(\psi) &= v(p_1 \psi) + v\left(\tau_{\pi}(p_2 \psi)\right) - \int_{\T} \nabla \widetilde{\phi}(x) \nabla\left(\frac{p_3 \psi}{\widetilde{\phi}}\right)(x) \mathrm{d} x \\
    &= v(p_1 \psi) + v\left(\tau_{\pi}(p_2 \psi)\right) - \int_{I_3} \frac{\nabla \widetilde{\phi}(x)}{\widetilde{\phi}(x)} \left(\nabla p_3(x) \psi(x) + p_3(x) \nabla \psi(x)\right) - \frac{\nabla \widetilde{\phi}(x)^2}{\widetilde{\phi}(x)^2} p_3(x) \psi(x) \mathrm{d} x, 
\end{align*}
where $(\tau_\pi \psi)(x) = \psi(x-\pi)$. As $p_j \in C^\infty(\T)$, the first and second term clearly define distributions in $\mH^{-1}(\T)$. Moreover, since $\widetilde{\phi} >c$ inside $I_3$, by the GNS inequality~\eqref{eq:GNS inequality}, we have
\begin{align*}
    |\widetilde{v}(\psi)| \lesssim \left(\norm{v}_{\mH^{-1}} + 1+ \frac{\norm{\widetilde{\phi}}_{\mH^1}^2}{c^2}\right)\norm{\psi}_{\mH^1(\T)}, \quad \mbox{for any $\psi \in \mH^1(\T)$.}
\end{align*}
Hence $\widetilde{v} \in \mH^{-1}(\T)$. We now observe that, since each $p_j$ is supported on $I_j$, from~\eqref{eq:local} and the ground-state identity 
\begin{align*}
    \inner{h(v) \phi_v, \psi}= \int_\T \nabla \phi_v(x) \nabla \psi(x) \mathrm{d} x + v(\phi_v \psi) = 0, \quad \mbox{for any $\psi \in \mH^1(\T)$,}
\end{align*}
we have
\begin{align*}
    -\Delta \widetilde{\phi}(p_1 \psi) &= \int_{I_1} \nabla \widetilde{\phi}(x) \nabla(p_1 \psi)(x) \mathrm{d} x = \int_{\T} \nabla \phi_v(x) \nabla(p_1 \psi)(x) \mathrm{d} x = -v(p_1 \phi_v \psi) = -v(p_1  \widetilde{\phi}\psi ), \quad \mbox{for any $\psi \in \mH^1(\T)$,}\\
    -\Delta \widetilde{\phi}(p_2 \psi) &= \int_{I_2} \nabla \widetilde{\phi}(x) \nabla (p_2 \psi)(x) \mathrm{d} x = \int_\T \nabla \phi_v(x-\pi) \nabla(p_2 \psi)(x) \mathrm{d} x = \int_\T \nabla \phi_v(x) \nabla(\tau_{\pi}p_2 \psi)(x) \\
    &= -v\left(\phi_v \tau_{\pi}( p_2 \psi)\right) = -v\left(\tau_{\pi} (p_2  \widetilde{\phi}\psi)\right), \quad \mbox{for any $\psi \in \mH^1(\T)$,}
    \intertext{and}
    -\Delta \widetilde{\phi}(p_3 \psi) &= \int_{I_3} \nabla \widetilde{\phi}(x) \nabla (p_3 \psi)(x) \mathrm{d} x = \int_\T \nabla \widetilde{\phi}(x)  \nabla \left(\frac{p_3}{\widetilde{\phi}} \widetilde{\phi} \psi\right)(x) \mathrm{d} x, \quad \mbox{for any $\psi \in \mH^1(\T)$.}
\end{align*}
Hence, by the partition of the unity property~\eqref{eq:partition unity} and the definition of $\widetilde{v}$ we have
\begin{align*}
    -\Delta \widetilde{\phi}(\psi) &= -\Delta \widetilde{\phi}\left(\sum_{j=1}^3 p_j \psi\right) = \sum_{j=1}^3 -\Delta \widetilde{\phi}(p_j \psi) \\
    &= -v(p_1 \widetilde{\phi}\psi) -v\left(\tau_{\pi} (p_2 \widetilde{\phi}\psi)\right) + \int_\T \nabla \widetilde{\phi}(x)  \nabla \left(\frac{p_3  \widetilde{\phi} \psi}{\widetilde{\phi}}\right)(x) \mathrm{d} x   = -\widetilde{v}( \widetilde{\phi} \psi), \quad \mbox{for any $\psi \in \mH^1(\T)$.}
\end{align*}
Therefore $\widetilde{\phi}$ is an eigenfunction of $h(\widetilde{v})$. Since both $\widetilde{\phi}$ and the ground-state of $h(\widetilde{v})$ are almost everywhere strictly positive (by Lemma~\ref{lem:non-degenerate}), they can not be orthogonal to each other. Therefore, by the non-degeneracy of the ground-state of $h(\widetilde{v})$ (see Lemma~\ref{lem:non-degenerate}), the function $\widetilde{\phi}$ must be the ground-state of $h(\widetilde{v})$, which yields the desired contradiction and concludes the proof.
\end{proof}

We can now prove Corollary~\ref{cor:necessary conditions}.
\begin{proof}[Proof of Corollary~\ref{cor:necessary conditions}] Let $v\in \mathcal{V}$ and denote by $\phi_v$ the ground-state of $h(v)$. Since this ground-state is non-degenerate, it follows from Lemma~\ref{lem:ground-state rep} that any normalized ground-state of $H_N(v)$ must have $\phi_v$ as a natural orbital with occupation number $1$. Thus, by Theorem~\ref{thm:non-degenerate} and~\eqref{eq:natural orbital decomposition} we have
\begin{align*}
    \rho_{\Psi}(x) = \gamma_{\Psi}(x,x) \geq |\phi_v(x)|^2 > 0 \quad \mbox{for any $x\in \T$.}
\end{align*}
Hence, any pure ground-state has nowhere vanishing density. As the density of a mixed ground-state is a convex combination of the density of pure ground-states, we conclude that any mixed ground-state has nowhere vanishing density. 

The integral constraint $\int \rho_{\Psi} = N$ comes from the normalization of $\Psi$ and the regularity condition $\sqrt{\rho_{\Psi}} \in \mH^1(\T)$ comes from the fact that $\Psi$ has finite kinetic energy (and follows from standard arguments, see, \cite{Lie83,SPR+24}). This completes the proof.
\end{proof}
\section{The non-interacting Hohenberg-Kohn theorem for distributional potentials}
\label{sec:HK}
We now turn to the proof of Theorems~\ref{thm:HK} and~\ref{thm:HK single-particle}. We begin with Theorem~\ref{thm:HK single-particle}.

\begin{proof}[Proof of Theorem~\ref{thm:HK single-particle}]
First, note that the map $v^{\rm KS}_1(\rho) = \frac{\Delta \sqrt{\rho}}{\sqrt{\rho}}$ should be understood in the sense 
\begin{align*}
    v^{\rm KS}_1(\rho)(\phi) = \frac{\Delta \sqrt{\rho}}{\sqrt{\rho}}(\phi) = \Delta \sqrt{\rho}\left(\frac{\phi}{\sqrt{\rho}}\right) =- \int_{\T} \nabla \sqrt{\rho}(x) \nabla \left(\frac{\phi}{\sqrt{\rho}}\right)(x) \mathrm{d} x = \Delta \sqrt{\rho} \left(M_{1/\sqrt{\rho}} \phi\right),
\end{align*}
where $M_{1/\sqrt{\rho}}$ is the multiplication operator 
\begin{align*}
    \phi \mapsto M_{1/\sqrt{\rho}}(\phi)= \frac{\phi}{\sqrt{\rho}}. 
\end{align*}
Since, for any $\rho\in \mathcal{D}_1$ we have $\rho \geq c$ for some $c>0$, we have $\sqrt{\rho} \in \mH^1(\T)$. Thus, from Lemma~\ref{lem:algebra}, the multiplication operator $M_{1/\sqrt{\rho}}$ is bounded in $\mH^1(\T)$ and therefore $v^{\rm KS}_1(\rho) \in \mathcal{V}$ for any $\rho \in \mathcal{D}_1$. Moreover, by construction we have
\begin{align*}
    -\Delta \sqrt{\rho}(\phi) + v(\sqrt{\rho} \phi) = 0, \quad \mbox{for any $\phi \in \mH^1(\T)$, where $v=v^{\rm KS}_1(\rho)$.}
\end{align*}
Therefore, $\sqrt{\rho}$ is an eigenfunction of $h(v)$. As $\sqrt{\rho}$ is strictly positive, it must be the ground-state of $h(v)$ by Theorem~\ref{thm:non-degenerate}. Hence, $v^{\rm KS}_1$ is indeed the Kohn-Sham map.

To prove that $\rho \mapsto v^{\rm KS}_1(\rho)$ is smooth, we shall show that it is a composition of smooth maps. So first, note that the map $\psi \in \mH^1(\T)\mapsto f(\psi) = \Delta \psi \in \mH^{-1}(\T)$ is linear and continuous, hence smooth. Moreover, in the proof of Lemma~\ref{lem:differentiable}, we have shown that the $M$ map $\psi \mapsto M_{\psi}$ is smooth from $\mH^1(\T)$ to $\mathcal{B}(\mH^1(\T))$. Furthermore, the map $h:\mH^{-1}(\T) \times \mathcal{B}(\mH^1(\T))\rightarrow \mH^{-1}(\T)$ given by
\begin{align*}
    (v, T) \rightarrow h(v,T) = v \circ T \in \mH^{-1}(\T) , \quad \mbox{where}\quad v\circ T(\phi) = v(T\phi), \quad\mbox{for $\phi \in \mH^1(\T)$,}
\end{align*}
is bilinear and continuous, hence, also smooth. Since
\begin{align*}
    v^{\rm KS}_1(\rho) = h\left(f(\sqrt{\rho}), M_{1/\sqrt{\rho}}\right),
\end{align*}
it suffices to show that the maps $\rho \in \mathcal{D}_1 \mapsto \sqrt{\rho}$ and $\rho \mapsto 1/\sqrt{\rho} \in \mathcal{D}_1$ are also smooth. Moreover, as $\mathcal{D}_1$ is a submanifold of the set $\mathcal{D}$ introduced in~\eqref{eq:D set} (see Remark~\ref{rem:Lie group}), we only need to show that these maps are smooth on $\mathcal{D}$. To this end, we observe that, since any $\rho\in \mathcal{D}$ satisfies $1/c\leq \rho \leq c$ for some $c = c(\rho)>0$, by the GNS inequality~\eqref{eq:GNS inequality} we can find a neighborhood $U_\rho\subset \mathcal{D}$ of $\rho\in \mathcal{D}$ such that $1/(2c(\rho)) \leq \rho' \leq 2 c(\rho)$ for any $\rho' \in U_\rho$. In particular, we can find $g_1,g_2\in C^\infty(\R;\R)$ such that $g_1(x) =\sqrt{x}$ and $g_2(x) = 1/\sqrt{x}$ for any $1/(2c(\rho)) \leq x \leq 2c(\rho)$. Therefore, by Lemma~\ref{lem:differentiable}, the maps $\rho' \in U_{\rho}\mapsto g_1(\rho')= \sqrt{\rho'}$ and $\rho'\in U_{\rho}\mapsto g_2(\rho') = 1/\sqrt{\rho'}$ are smooth. As smoothness is a local property, this shows that $\rho \mapsto v^{\rm KS}_1(\rho)$ is smooth.

To complete the proof, we need to show that $v^{\rm KS}_1(\rho)$ is the unique potential generating $\rho$. So suppose that $\rho \in \mathcal{D}_1$ is the ground-state density of $h(v)$ and $h(v')$ for $v,v'\in \mH^{-1}(\T)$. Without loss of generality, we also assume that both ground-state energies are $0$. Since the ground-state is strictly positive and non-degenerate by Theorem~\ref{thm:non-degenerate}, both operators have the same ground-state wave-function $\phi = \sqrt{\rho} \in \mH^1(\T)$. Hence, we have
\begin{align}
    0 = \inner{\phi, h(v) \psi}- \inner{\phi, h(v') \psi} = (v-v')(\overline{\phi} \psi), \quad\mbox{for any $\psi \in \mH^1(\T)$.} \label{eq:HKid}
\end{align}
As $\phi$ is strictly positive, the operator of multiplication by $\overline{\phi}$ is an isomorphism in $\mH^1(\T)$ by Lemma~\ref{lem:algebra}. Equation~\eqref{eq:HKid} thus implies that $(v-v') = (v-v')\circ M_{\overline{\phi}} \circ M_{1/\overline{\phi}} = 0$, which concludes the proof.
\end{proof}

We can now proceed to the proof of Theorem~\ref{thm:HK}.
\begin{proof}[Proof of Theorem~\ref{thm:HK}]
Let $v, v'\in \mathcal{V}$ and let $N\in \N$. Suppose that $\Psi$ and $\Psi'$ are ground-state wave-functions of $H_N(v)$ and $H_N(v')$ with the same density $\rho \in \mathcal{D}_N$. Moreover, without loss of generality, let us assume that both ground-state energies are zero. We now use the standard Hohenberg-Kohn argument to show that $\Psi$ is also a ground-state of $H_N(v')$. Precisely, we note that, since $\Psi$ is a ground-state of $H_N(v)$ we have 
\begin{align*}
    0 = \inner{\Psi' , H_N(v') \Psi'} \leq \inner{\Psi, H_N(v') \Psi} = \inner{\Psi, H_N(v) \Psi} + \left(v'-v\right)(\rho) = (v'-v)(\rho). 
\end{align*}
As the same argument holds when we exchange $v$ and $v'$, we conclude that $(v'-v)(\rho) = 0$. In particular, the above is an equality and $\Psi$ is also a ground-state of $H_N(v')$. Moreover, we note that, if $\rho$ is only representable by a mixed ground-state, i.e., $\rho = \sum_{j} \lambda_j \rho_{\Psi_j}$ for some $\lambda_j \geq 0$ and $\Psi_j$ ground-states of $H_N(v)$, the same argument shows that each of the $\Psi_j$ is a mutual ground-state of $H_N(v)$ and $H_N(v')$ (see, e.g., \cite[Theorem 1]{PTC+23} for more details). So without loss of generality, we denote by $\Psi$ (one of) the mutual ground-states of $H_N(v)$ and $H_N(v')$.

Since $\Psi$ is a simultaneous ground-state of both $H_N(v')$ and $H_N(v)$, we must have
\begin{align*}
    \inner{\left(H_N(v)-H_N(v')\right)\Psi,  \Phi} = (v-v')(\rho_{\Psi\Phi}) =0, \quad \mbox{for any $\Phi \in \mathcal{H}_N \cap \mH^1(\T^N)$,}
\end{align*}
where $\rho_{\Psi\Phi}$ is the overlapping density
\begin{align*}
    \rho_{\Psi \Phi}(x) = N \int_{\T^{N-1}} \overline{\Psi(x,x_2,...,x_N)} \Phi(x,x_2,...,x_N) \mathrm{d} x_2...\mathrm{d} x_N.
\end{align*}
As $\Psi$ is the ground-state of $H_N(v)$ and $h(v)$ has discrete spectrum, by Lemma~\ref{lem:ground-state rep} we can choose a basis of eigenfunctions $\{\phi_j\}_{j \in \N}$ such that $\{\phi_j\}_{j\leq M}$ for some $M<\infty$ are the natural orbitals (with non-zero occupation number) of $\Psi$. Thus, if we choose trial states of the form $\Phi = a(\phi_\ell)^\ast a(\phi_k) \Psi$ for $k\leq M$ and $\ell \geq M+1$, from the formula in Lemma~\ref{lem:overlap}  we have
\begin{align}
    n_k (v-v')(\overline{\phi}_k \phi_\ell) = 0, \quad \quad \mbox{for any $\ell \geq M+1$ and $1\leq k \leq M$} \label{eq:HKeq}
\end{align}

We can now use the representation in Lemma~\ref{lem:dual rep} to show that $(v-v')$ is more regular than expected. Precisely, note that the operator $M_{\overline{\phi_k}}$ of multiplication by $\overline{\phi_k}$ is a bounded operator in $\mH^1(\T)$ (see Lemma~\ref{lem:algebra}). Therefore, it follows from~\eqref{eq:HKeq}, the fact that $n_j >0$ for any $j\leq M$, and the representation in Lemma~\ref{lem:dual rep} that
\begin{align}
    (v-v')\circ M_{\overline{\phi_k}} (\psi) = \sum_{j=1}^M A_{kj} \inner{\phi_j, \psi} \quad \mbox{for some $\{A_{k j}\}_{k,j \leq M} \in \C^{M\times M}$ and $\psi \in \mH^1(\T)$. } \label{eq:HKeq2}
\end{align}
As the ground-state $\phi_1$ is strictly positive by Theorem~\ref{thm:non-degenerate}, the operator $M_{1/\phi_1}$ is bounded in $\mH^1(\T)$ by Lemma~\ref{lem:algebra}; hence,
\begin{align*}
    (v-v')(\psi) = (v-v')\circ M_{\overline{\phi_1}}\circ  M_{1/\overline{\phi_1}}(\psi) = \sum_{j=1}^M A_{1j} \left\langle \frac{\phi_j}{\phi_1},\psi\right\rangle, \quad \mbox{for any $\psi \in \mH^1(\T)$.}
\end{align*}
In other words, the distribution $(v-v')$ can be identified (via the Riesz map~\eqref{eq:Riesz}) with the function
\begin{align}
    (v-v')(x) = \sum_{j=1}^M A_{1j} \frac{\phi_j(x)}{\phi_1(x)}. \label{eq:v function}
\end{align}
In particular $(v-v')\in \mH^1(\T)$. Hence, to conclude the proof, it suffices to show that $(v-v')$ is piecewise constant. Indeed, if $(v-v')$ was piecewise constant but not constant, it would have a jump, which is not possible for $\mH^1$ functions. 

To see that $(v-v')$ is piecewise constant, note that by~\eqref{eq:HKeq2} we have
\begin{align*}
    (v-v')(x) \frac{\phi_k(x)}{\phi_1(x)} = \sum_{j=1}^N A_{kj} \frac{\phi_j(x)}{\phi_1(x)}, \quad \mbox{for any $1\leq k \leq M$ and (almost) every $x\in \T$.}
\end{align*}
In other words, the above equation shows that the vector
\begin{align*}
    \vec{\Phi}(x) \coloneqq \left(1 , \frac{\phi_2(x)}{\phi_1(x)},...,\frac{\phi_M(x)}{\phi_1(x)}\right)^T \neq 0
\end{align*}
is an eigenvector of the matrix $A=\{A_{kj}\}\in \C^{M \times M}$ with corresponding eigenvalue $(v-v')(x)$ for (almost) every $x\in \T$. As $A\in \C^{M\times M}$ can have at most $M$ distinct eigenvalues, we conclude that $(v-v')(x)$ is piecewise constant, which completes the proof.
\end{proof}


We now prove Theorem~\ref{thm:excited}.
\begin{proof}[Proof of Theorem~\ref{thm:excited}] Let $\phi_k$ be a real-valued excited state of $h(v)$ for some $v\in \mH^{-1}(\T)$. Since the ground-state $\phi_1$ is non-degenerate and strictly positive by Theorem~\ref{thm:non-degenerate}, there must exist some $x_0 \in \T$ such that $\phi_k(x_0) = 0$, as otherwise $\phi_k$ would not change sign and the overlap $\inner{\phi_k, \phi_1}$ would be non-zero, contradicting the orthogonality with $\phi_1$. We can now consider perturbations of $v$ of the form $v_\alpha = v+ \alpha \delta_{x_0}$ for $\alpha \in \R$. Indeed, as $\phi_k$ is an eigenfunction of $h(v)$ with $\phi_k(x_0)=0$, it follows that
\begin{align*}
    q_{h(v)}(\phi_k, \psi) = \inner{h(v) \phi_k, \psi} =  \inner{h(v) \phi_k,\psi} + \alpha \overline{\phi_k(x_0)} \psi(x_0) = \inner{h(v_\alpha) \phi_k,\psi},\quad \mbox{for any $\psi \in \mH^1(\T)$.}
\end{align*}
Thus $\phi_k$ is also an excited state of $h(v_\alpha)$ for any $\alpha \in \R$. Moreover, as the spectrum of $h(v)$ is discrete, for $\alpha>0$ small enough, $\phi_k$ must be the $k^{th}$ excited state\footnote{Assuming $k-1\in \N$ is the previous closed shell in the case of degeneracies, i.e., $\lambda_{k-1} < \lambda_k$. } of $h(v_\alpha)$ by standard perturbation theory.
\end{proof}

\section{Concluding remarks}
\label{sec:conclusion}
In this paper, we obtained a complete characterization of the set of densities that are representable by non-interacting Schr\"odinger operators with a certain class of distributional potentials in the one-dimensional torus. Moreover, we proved a Hohenberg-Kohn theorem for such operators, thereby establishing the uniqueness of the Kohn-Sham density-to-potential map and, thanks to \cite[Corollary 19]{SPR+24}, the differentiability of the non-interacting convex Lieb functional. In particular, our results show that, in the non-interacting case, the class of distributions in $\mH^{-1}(\T)$ is not only sufficient but also necessary to represent all strictly positive densities coming from wave-functions with finite kinetic energy. 

Let us now comment on possible extensions of these results and further open questions.\begin{enumerate}[label=(\roman*)]
    \item First, we emphasize that no necessary conditions for \emph{interacting} $\mathcal{V}$-representability were established here. Thus an immediate open question is how to extend Theorems~\ref{thm:representability} and~\ref{thm:HK} to the case of interacting systems on $\T$. 
    \item Second, we note that the unique continuation property restricted to ground-states in Lemma~\ref{lem:non-degenerate} relies on powerful abstract results of Reed and Simon that can be applied to a much more general setting such as higher-dimensional Schr\"odinger operators. While these results are not applicable to interacting systems due to Fermi statistics (see the comments after  \cite[Theorem X.III.46]{RS78}), they can still be used to establish a Hohenberg-Kohn theorem for non-interacting systems, and therefore, the uniqueness of the Kohn-Sham density to potential map in a much broader context (e.g., for bosons).
    \item Third, we note that Theorems~\ref{thm:non-degenerate} and~\ref{thm:HK} can be extended to the case of an interval $I=[0,2\pi]$ with Dirichlet boundary conditions. However, it is not true that all densities that are strictly positive inside $I$, and come from wave-functions with finite kinetic energy, are representable by potentials in the dual space of $\mH^1_0(I)$. This can be seen, for instance, by considering densities like $\sqrt{\rho}(x) = x^2$ for $x$ close to $0$, in the case of a single-particle. Indeed, in this case, $\Delta \sqrt{\rho}/\sqrt{\rho}(x) = 2/x^2$, which does not define a continuous functional in $\mH^1_0(I)$. Therefore, it would be interesting to systematically study the properties of the single-particle Kohn-Sham map~\eqref{eq:KSmap} for functions in $\mH^1_0(I)$. This could lead to further insights into natural necessary and sufficient conditions for $v$-representability in non-compact spaces such as the line $\R$, or even to higher dimensional spaces.
\end{enumerate}
\addtocontents{toc}{\protect\setcounter{tocdepth}{1}}

\appendix
\addtocontents{toc}{\protect\setcounter{tocdepth}{-1}}
\section*{Acknowledgements}
The author is grateful to Markus Penz, Sarina Sutter, and Michael Ruggenthaler for helpful discussions, useful remarks on the first draft of this paper, and inspiring presentations during the workshop on Foundations and Extensions of DFT in Oslo. The author is also grateful to Andre Laestadius and Vebjørn Bakkestuen for organizing the aforementioned workshop which led to the ideas for the current paper. Special thanks also to Asbjørn Bækgaard Lauritsen for calling attention to Remark~\ref{rem:natural orbital} and spotting a mistake in the previous proof of Theorem~\ref{thm:HK}. 

T.C.~Corso acknowledges funding by the \emph{Deutsche Forschungsgemeinschaft} (DFG, German Research Foundation) - Project number 442047500 through the Collaborative Research Center "Sparsity and Singular Structures" (SFB 1481). 

\addtocontents{toc}{\protect\setcounter{tocdepth}{2}}




\section*{Data availability}
No datasets were generated or analysed during the current study.

\section*{Competing interests}

The authors have no competing interests to declare that are relevant to the content of this article.

\bigskip

\begin{thebibliography}{EFGHW21}


\bibitem[AS88]{AS88}
    \textsc{F.~Aryasetiawan} and \textsc{J.~M.~Stott},
    \newblock Effective potentials in density-functional theory,
    \newblock \doi{10.1103/PhysRevB.38.2974}{\emph{Physical Review B}} \textbf{38} (1988), no.5, pp. 2974--2987.
    \hfill


\bibitem[CCR85]{CCR85}
    \textsc{J.~T.~Chayes},  \textsc{L.~Chayes}, and \textsc{M.~B.~Ruskai},
    \newblock Density functional approach to quantum lattice systems,
    \newblock \doi{10.1007/BF01010474}{\emph{Journal of Statistical Physics}} \textbf{38} (1985), no.3--4, pp. 497--518.
    \newblock \mr{788430}.
    \newblock \zbl{0632.46072}.
    \hfill

\bibitem[CH89]{CH89}
    \textsc{R.~Courant} and \textsc{D.~Hilbert},
    \newblock \doi{10.1002/9783527617210}{\emph{Methods of mathematical physics. {V}olume {I}}}, John Wiley \& Sons, (1989).
    \mr{65391}.
    \zbl{0729.00007}.
  
\bibitem[CS91]{CS91}
    \textsc{J.~Chen} and \textsc{M.~J.~Stott}, 
    \newblock v-representability for systems with low degeneracy,
    \newblock \doi{10.1103/PhysRevA.44.2816}{\emph{Physical Review A}} \textbf{44}, no. 5, pp. 2816--2822.
    \hfill

\bibitem[CS93]{CS93}
    \textsc{J.~Chen} and \textsc{M.~J.~Stott},
    \newblock v-representability for noninteracting systems,
    \newblock \doi{10.1103/PhysRevA.47.153}{\emph{Physical Review A}} \textbf{47} (1993), no. 1, pp. 153--160.

\bibitem[Gar18]{Gar18}
    \textsc{L.~Garrigue},
    \newblock Unique continuation for many body {S}chr\"{o}dinger operators and the {H}ohenberg-{K}ohn theorem, \newblock\doi{10.1007/s11040-018-9287-z}{\emph{Mathematical Physics, Analysis and Geometry}} \textbf{21} (2018), no.3, Paper No. 27, 11.
    \newblock \mr{3854406}.
    \newblock \zbl{1400.81217}.
    \hfill

\bibitem[Gar19]{Gar19}
    \textsc{L.~Garrigue},
    \newblock Hohenberg-{K}ohn theorems for interactions, spin and temperature,
    \newblock \doi{10.1007/s10955-019-02365-6}{\emph{Journal of Statistical Physics}} \textbf{177} (2019), no.3, pp. 415--437.
    \newblock \mr{4026653}.
    \newblock \zbl{1429.81057}.
    \hfill

\bibitem[Geo79]{Geo79}
    \textsc{V.~Georgescu},
    \newblock On the unique continuation property for {S}chr\"{o}dinger {H}amiltonians,
    \newblock \emph{Helvetica Physica Acta} \textbf{52} (1979), no,5-6, pp. 655--670.
    \newblock \mr{576454}.
    \hfill

\bibitem[Her89]{Her89}
    \textsc{J.~Herczy\'nski, Jan},
    \newblock On {S}chr\"{o}dinger operators with distributional potentials,
    \newblock \doi{}{\emph{Journal of Operator Theory}} \textbf{21} (1989), no.2, pp. 273--295.
    \newblock \mr{1023316}.
    \hfill
    
\bibitem[HK64]{HK64}
    \textsc{P.~Hohenberg} and \textsc{W.~Kohn},
    \newblock Inhomogeneous electron gas, \doi{10.1103/PhysRev.136.B864}{\emph{Physical Review. Series II}} \textbf{136} (1964), B864--B871.
    \newblock \mr{180312}.
    \newblock \zbl{}.
    \hfill

\bibitem[KM97]{KM97}
    \textsc{A.~Kriegl} and \textsc{P.~W.~Michor},
    \newblock \doi{10.1090/surv/053}{\emph{The convenient setting of global analysis}}, American Mathematical Society (1997).
    \newblock \mr{1471480}.
    \newblock \zbl{0889.58001}.
    \hfill
    
\bibitem[Koh83]{Koh83}
    \textsc{W.~Kohn},
    \newblock $v$-Representability and Density Functional Theory,
    \newblock \doi{10.1103/PhysRevLett.51.1596}{\emph{Physical Review Letters}} \textbf{51} (1983), no. 51, pp. 1596--1598.
    \hfill
    

\bibitem[KS65]{KS65}
    \textsc{W.~Kohn} and \textsc{L.~J.~Sham},
    \newblock Self-consistent equations including exchange and correlation effects,
    \newblock \doi{10.1103/PhysRev.140.A1133}{\emph{Physical Review. Series II}} \textbf{140} (1965), A1133--A1138.
    \newblock \mr{189732}.
    \hfill

\bibitem[Kur97]{Kur97}
    \textsc{K.~Kurata},
    \newblock A unique continuation theorem for the {S}chr\"{o}dinger equation with singular magnetic field, 
    \newblock \doi{10.1090/S0002-9939-97-03672-1}{\emph{Proceedings of the American Mathematical Society}} \textbf{125} (1997), no.3, pp. 853--860.
    \newblock \mr{1363173}.
    \newblock \zbl{0887.35026}
    \hfill
    
\bibitem[Lam18]{Lam18}
    \textsc{P.~E.~Lammert},
    \newblock In search of the {H}ohenberg-{K}ohn theorem,
    \newblock \doi{10.1063/1.5034215}{\emph{Journal of Mathematical Physics}} \textbf{59} (2018), no.4, pp.042110, 19.
    \newblock \mr{3789875}.
    \zbl{1386.81157}.
    \hfill
    
\bibitem[LBP20]{LBP20}
    \textsc{A.~Laestadius}, \textsc{M.~Benedicks},  and \textsc{M.~Penz},
    \newblock Unique continuation for the magnetic Schrödinger equation,
    \newblock \doi{10.1002/qua.26149}{\emph{International Journal of Quantum Chemistry}} \textbf{120} (2020).

    
\bibitem[Lie83]{Lie83}
    \textsc{E.~H.~Lieb},
    \newblock Density functionals for Coulomb systems, 
    \newblock \doi{10.1002/qua.560240302}{\emph{International Journal of Quantum Chemistry}} \textbf{24} (1983), pp. 243–277.
    \hfill

\bibitem[PL21]{PL21}
    \textsc{M.~Penz}, and \textsc{R.~v.~Leeuwen},
    \newblock Density-functional theory on graphs,
    \newblock \doi{10.1063/5.0074249}{\emph{The Journal of Chemical Physics}} \textbf{155} (2021), no. 24, pp. 244111.
    \hfill



\bibitem[PTC+23]{PTC+23}
    \textsc{M.~Penz}, \textsc{E.~I.~Tellgren}, \textsc{M.~A.~Csirik}, \textsc{M.~Ruggenthaler},
    and \textsc{A.~Laestadius},
    \newblock The Structure of Density-Potential Mapping. Part I: Standard Density-Functional Theory,
    \newblock \doi{10.1021/acsphyschemau.2c00069}{\emph{American Chemical Society}} (2023), 3(4), 334-347.
    \hfill

\bibitem[Reg01]{Reg01}
    \textsc{R.~Regbaoui},
    \newblock Unique continuation from sets of positive measure,
    \newblock \doi{10.1007/978-1-4612-0203-5_13}{\emph{In: Carleman estimates and applications to uniqueness and control
    theory}} \textbf{46} (2001), 179--190.
    \newblock \mr{1839175}
    \newblock \zbl{1165.35332}.
    \hfill

\bibitem[RS75]{RS75}
    \textsc{M.~Reed} and \textsc{B.~Simon},
    \newblock \emph{Methods of modern mathematical physics. {II}. {F}ourier analysis, self-adjointness}, Academic Press (1975).
    \newblock \mr{493420}.
    \hfill

\bibitem[RS78]{RS78}
    \textsc{M.~Reed} and \textsc{B.~Simon},
    \newblock \emph{Methods of modern mathematical physics. {IV}. {A}nalysis of operators}, Academic Press (1978).
    \newblock \mr{493421}.
    \hfill

\bibitem[RS80]{RS80}
    \textsc{M.~Reed} and \textsc{B.~Simon},
    \newblock \emph{Methods of modern mathematical physics. {I}. Functional Analysis}, Academic Press (1980).
    \newblock \mr{751959}
    \hfill

\bibitem[SPR+24]{SPR+24}
    \textsc{S.~M.~Sutter}, \textsc{M.~Penz}, \textsc{M.~Ruggenthaler}, \textsc{R.~van Leuween}, and \textsc{K.~J.~H.~Giesbertz},
    \newblock Solution of the {$v$}-representability problem on a one-dimensional torus, 
    \newblock \doi{10.1088/1751-8121/ad8a2a}{\emph{Journal of Physics. A. Mathematical and Theoretical}} \textbf{57} (2024), no 47.
    \newblock \mr{4840211}.
    \newblock \zbl{07945786}.
    \hfill

\bibitem[SS80]{SS80}
    \textsc{M.~Schechter} and \textsc{B.~Simon},
    \newblock Unique continuation for {S}chr\"{o}dinger operators with unbounded potentials,
    \newblock \doi{10.1016/0022-247X(80)90242-5}{\emph{Journal of Mathematical Analysis and Applications}} \textbf{77} (1980), no.2, pp. 482--492.
    \newblock \mr{593229}.
    \newblock \zbl{0458.35024}.
    \hfill

\bibitem[THS+22]{THS+22}
    \textsc{A.~M.~Teale}, 
    \textsc{T.~Helgaker}, \textsc{A.~Savin}, \textsc{C.~Adamo}, \textsc{B.~Aradi}, \textsc{et al},
    \newblock DFT exchange: sharing perspectives on the workhorse of quantum chemistry and materials science
    \newblock \doi{10.1039/D2CP02827A}{\emph{Physical chemistry chemical physics}} \textbf{24} (2022), no. 47, pp. 28700-28781.
    \hfill
    
\bibitem[WAR+23]{WAR+23}
    \textsc{J.~Wrighton}, \textsc{A.~Albavera-Mata}, \textsc{H.~F.~Rodr\'iguez}, \textsc{T.~S.~Tan}, \textsc{A.~C.~Cancio}, \textsc{J.~W.~Dufty}, and \textsc{S.~B.~Trickey},
    \newblock Some problems in density functional theory,
    \newblock \doi{10.1007/s11005-023-01649-z}{\emph{Letters in Mathematical Physics}} \textbf{113} (2023), no.2.
    \mr{4572242}.
    \zbl{1515.81243}.
    \hfill

\bibitem[Zho12]{Zho12}
    \textsc{A.~Zhou},
    \newblock Hohenberg-{K}ohn theorem for {C}oulomb type systems and its generalization,
    \newblock \doi{10.1007/s10910-012-0061-3}{\emph{Journal of Mathematical Chemistry}} \textbf{50} (2012), no.10, pp. 2746--2754.
    \newblock \mr{2989094}.
    \newblock \zbl{1308.81197}.
    \hfill


\end{thebibliography}
\end{document}